# Overlay Alignment Using Two Photonic Crystals

Can Peng, Keith Morton, Zhaoning Yu, and Stephen Y. Chou
Nanostructure Laboratory, Department of Electrical Engineering
Princeton University, Princeton, NJ 08544

**Abstract**

**In this paper we proposed a novel overlay alignment method using two sets of identical photonic crystals (PhCs). In this method the reflection or transmission spectrum of the two overlaid photonic crystals is measured to help wafer tilt, yaw rotation, and translation aligning. The initial testing results with two 1D photonic crystals and analysis of the alignment accuracy are presented. This method is particularly useful in building photonic crystal stacks with nanoimprint lithography (NIL).**

# 1. Introduction

Accurate overlay alignment between two layers of patterns is extremely important in lithography. Traditional alignment methods include direct comparison of two sets of marks, Morie patterns of two sets of gratings with slightly different periods [1], and diffractions of two sets of gratings with an identical period. These traditional alignment methods do not provide high accuracy tilt alignment between two wafers that is very important to NIL. Here a novel overlay alignment method with new principle is proposed, which applies the resonance phenomenon in 2D PhCs [2]. This resonance phenomenon in one 2D PhCs leads to narrow band filtering in reflection [3]. And the resonating frequency is very sensitive to the incidental angle and polarization of the incidental light with proper design. When two PhCs are overlaid the transmission or reflection spectra can be used to do tilt, rotation alignment and even translation alignment if these two PhCs are close enough to each other.

The rest of this paper is organized as following: in section 2 the basic principle of this method is presented and the alignment accuracy is analyzed, the experimental results are presented in section 3, finally conclusion is given in section 4.

# 2. Theory

The principle of this new method is based on the resonance phenomenon in planar periodic structures caused by coupling between incidental light and guided modes in the structures []. For simplicity only 1D structures are considered here. Figure 1 shows the structure of 1D PhC slabs. A silicon nitride layer are deposited on glass substrate and a lay of PhC slab is fabricated on the top by nanoimprint lithography. If the top PhC is much thinner than the wavelength of incidental light, the reflection spectra of this kind of resonant periodic structures around the resonating wavelength can be approximated by Lorentzian lineshape expressed as []:

$$R = \frac{|c|^2}{(k_{zm} - \beta_r) + \beta_i^2} \quad (1),$$

where

$$k_{zm} = |\vec{k}_z - m\vec{K}| \quad (2).$$

$K$ is the grating constant, $\vec{k}_z$ is the in-plane component of wavevector of the incidental light. $m$ is an integer. $\beta_r$ and $\beta_i$ are respectively real part and image part of the propagating constant of guided wave in the slab waveguide formed by the SiNx layer (the image part is introduced by the top grating). So the resonating condition is the quasi momentum conservation along the grating plane. For 1D PhCs two kinds of incidental angles are considered, one is defined as tilt angle (Fig.1a) and another is defined as yaw (Fig.1b).

The resonating conditions for these two situations are given by

$$k \sin\theta_{tilt} \pm mK = \pm\beta_r(k) \quad (3),$$

$$\sqrt{k^2 \sin^2\theta_{yaw} + (mK)^2} = \beta_r(k) \quad (4),$$

where $k$ is the wavevector of incidental light in free space, and for subwavelength cases m always equals one. From equation (3) it is easy to be seen that if tilt angle is not zero coupling conditions for the incidental light to the guided wave in two directions are different. This causes two minimum transmission wavelengths in spectrum. The yaw angle causes blue shift of minimum transmission position according to equation (4). The effects of tilt angle on minimum transmission splits and effects of yaw angle on minimum transmission shift can be calculated numerically with rigorous coupled-wave analysis (RCWA) []. In the proposed novel alignment method the splits and blue shift of the minimum transmissions can be used to adjust the parallelity of the two overlaid PhCs.

If the two overlaid PhCs are face-to-face and parallel to each other and the gratings are also parallel to each other, as shown in principle scheme (Fig.2), translation between of the two PhCs along the direction perpendicular to the grating can affect the transmission spectrum. When the gap between the two PhCs is small enough, the waveguided modes of the two PhCs are coupled with each other. And the transmission spectrum of the photonic crystal stack is sensitive to the translation shift between them. The numerical investigation results of the effects of translational shift are shown in Fig.3. In simulation UV curable resist is assumed to fill the gap between the two PhCs, which is most of cases in alignment processes. From the simulation results it can be seen that the spectrum for zero shift between two PhCs and that for reverse phase shift (half pitch shift) has single minimum transmission in considered wavelength range, other values of shift cause double minima of transmission. This phenomenon can be explained by zero-pole theory of gratings. The overlaid two PhCs can be viewed as a system with two poles and one zero, and the transmission spectrum of this system can be expressed in such a formation

$$T = 1 - A \frac{k - z_1}{(k - p_1)(k - p_2)} \quad (5)$$

The positions of the poles and the zero in complex plane depend on the relatively translation shift between the two overlaid PhCs. And it can be proved that the zero coincides with one of the two poles when the translation shift between the two PhCs is 0 or half pitch (this will be presented in another work). That means for these two kinds of shift the whole structure is a one-pole system and has only one transmission minimum in certain range. The accuracy of the translational alignment with this method depends on the resolution of the spectrometer used. For the structure used in the simulation, the translational alignment accuracy can be estimated as 20nm when spectrometer with 0.01nm resolution is used.

Because the coupling resonance between incidental light and guided modes in PhCs depends on the polarization of the incidental light, special 2D photonic crystal structures may be applied to do rotation alignment (azimuthal alignment, considering the direction of incidental light as axis). That would not be discussed in this paper.

In principle transmission spectrum or reflection spectrum of two overlaid special PhCs (with in-plane waveguiding ability) can be used to do tilt, translation and rotation alignment. Next section preliminary experiments are presented and the alignment accuracy is discussed.

## 3. Experiments and discussions

The scheme of the experimental setup is shown in Fig.4. The light source is a light emission diode (LED) with central wavelength of 1537nm and full width of half maximum (FWHM) of 49.4nm. The light is coupled out with a collimator from fiber to free space and goes through a polarizer to be linearly polarized. This light then goes through two samples of photonic crystals held by stages that can adjust the tilt angle and yaw angle of samples. The transmitted light is collected by another collimator and sent to optical spectrum analyzer. The PhC samples used in experiment are 1 um 1D gratings. The structure of these samples is similar as that shown in Fig.3(c), but the parameters are different. In these samples the silicon nitride layer under the grating is 350 nm thick and the grating depth is 49 nm.

The results of experiments of tilt and yaw alignments are show in Fig.5. Because the two PhC samples are assumed to be the same in structure, when they are exactly parallel to each other and the light normally incidents the transmission spectrum has just one minimum and with a narrow linewidth. If there is a tilt angle between these two samples one of these samples is not perpendicular to the incidental light beam any more and there are three minima of transmission in the spectrum (as shown in Fig.5 (a)). The outer two minima of transmission are caused by the splitting of transmission minimum of the tilted sample. The splitting between two outer minima was measured under different tilt angle and the result is shown in Fig.5 (b). The result shows that the variation of transmission spectrum is very sensitive to the tilt angle. The data was fitted by a linear function showing that 0.03 degree tilt angle causes 1 nm split of the transmission minima. Then with sub-angstrom spectrometer the accuracy of tilt angle alignment can reach the order of 0.001 degree.

The variation of spectrum versus yaw angle (defined in Fig.1(b)) was also measured. The results are shown in Fig.6. From equation (4) it can be seen that yaw angle causes blueshift of minimum in transmission spectrum of the yawed sample. The results of measurement confirmed this as shown in Fig.6(a). The blueshift versus yaw angle is shown in Fig.6(b). The effect of yaw angle on transmission spectrum is much smaller than that of tilt angle. Measurement data show that 0.5 degree yaw angle can only cause 1 nm blueshift of transmission minimum.

The much higher sensitivity of transmission to tilt angle than to yaw angle suggests that 2D photonic crystal structure is more suitable in this application, because for 2D

PhCs the effects of tilt and yaw on the transmission spectra both can be describe by equation (3) then lead to high alignment accuracy. Proper choose of material and parameters of the structure can improve the alignment accuracy further. From equation (3) it can be seen that the smaller the chromatic dispersion the higher the sensitivity of transmission spectrum to the incidental angle. So waveguiding structure with low waveguiding chromatic dispersion is preferred in this alignment application. The relationship between the split of transmission minima and small tilt angle can be expressed as

$$\Delta\lambda \approx \frac{4\pi\theta_{tilt}}{\lambda_0 \frac{\partial \beta_r}{\partial \lambda_0}} \qquad (5)$$

It is easy to be drawn out that thick slab waveguides have small waveguiding chromatic dispersion. Several things should be considered in designing the structure, not only small chromatic dispersion is pursued, but also the position of minimum transmission and linewidth of the transmission valley should be considered. The position of the minima needs to be in the spetrum range of light source, and the linewidth cannot be too small to be resolved by optical spectrum analyzers.

## 4. Conclusions

A novel optical method for overlay alignment using two PhCs is proposed in this work. The mechanism of this method is based on the coupling between the incidental light and the in-plane guided wave of the PhCs through the periodic structure. This coupling leads to resonance for some certain wavelengths and then minima transmission in spectra. Theoretical and experimental investigations show that the transmission spectra are sensitive to the incidental angle the polarization of light and relative translation between the two overlaid PhCs. With this method, tilt, rotation and translation alignments can be achieved. The accuracy of alignments mainly depends on the structure of PhCs and the resolution of the optical spectrum analyzer used to detect the transmitted light.

This new method is very useful for nanoimprint lithography (NIL) and using NIL to build photonic crystal stacks and it's also a new method of postural controlling for other applications.

Fig.1 Definition of tilt and yaw and illustration of resonance conditions for these two kinds of incidental angles and their effects on transmission spectrum.

Fig.2 Principle scheme for alignment with two PhCs. When the two periodic structures are in phase or have $\pi$ phase shift, the system is symmetric in horizontal direction. In other phase shift it is asymmetric.

Fig.3 The simulated transmission spectra of two overlaid PhCs (UV resist in between them, polarization of incidental light is perpendicular to the grating), (a) the spectra when the gap between two PhCs is 600nm, (b) the gap is 1000nm. The curves labeled A-F are corresponding to situations of 0um, 0.1um, 0.2um, 0.3um, 0.4um and 0.5um shift between two PhCs respectively. The structure of the PhCs is shown in (c), the duty cycle is 0.5.

Fig.4 Scheme of experimental setup

Fig.5. (a) some transmission spectra under different tilt angles. And the illustration of tilt is shown at the right bottom of (a). (b) the measurement of the relationship between the split of transmission minimum and the tilt angle, and the linear fit of the data.

Fig.6. (a) transmission spectra under different yaw angles and schematic illustration of definition of yaw angle. (b) relationship between blueshift of transmission minimum and the yaw angle.

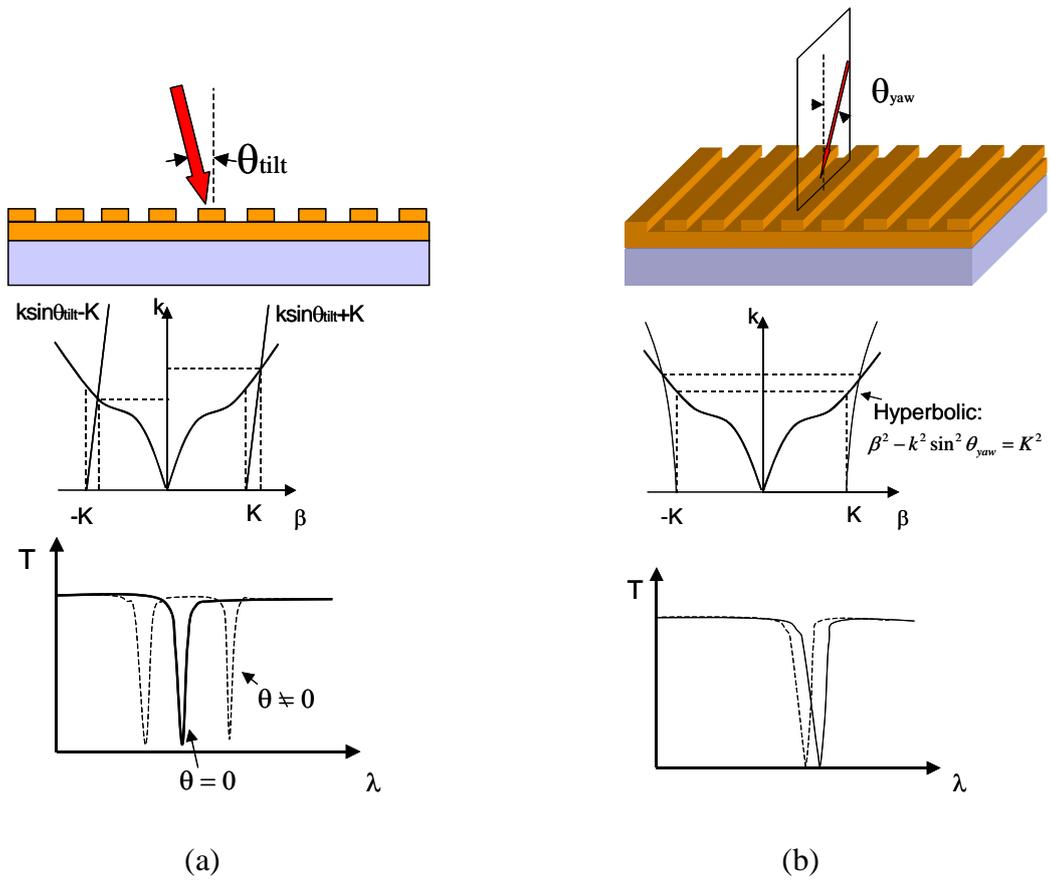

(a)                      (b)

Fig.1

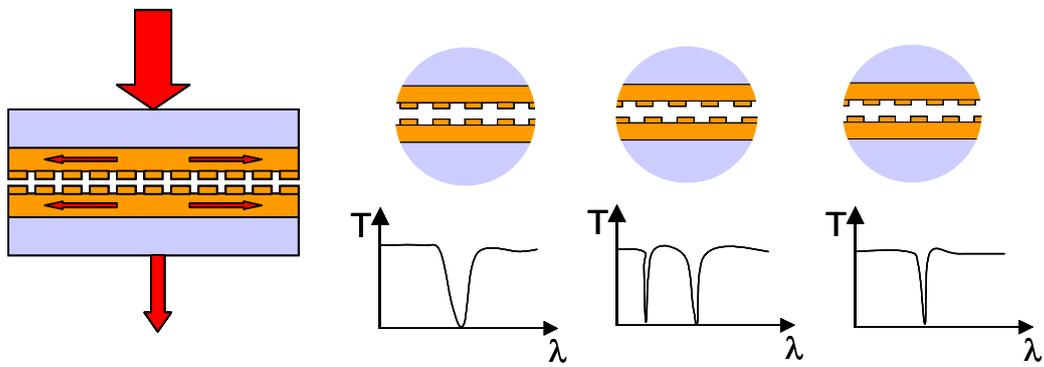

Fig.2

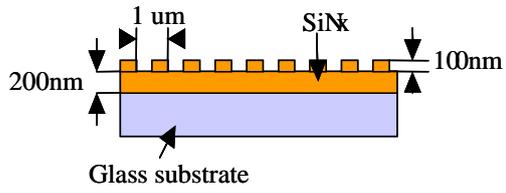
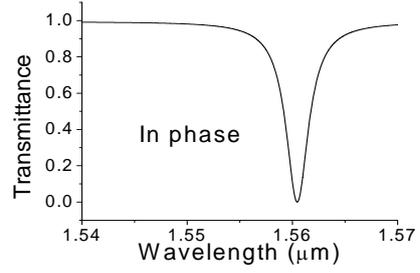
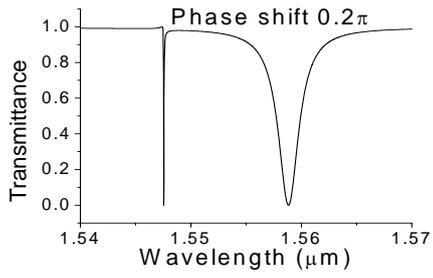
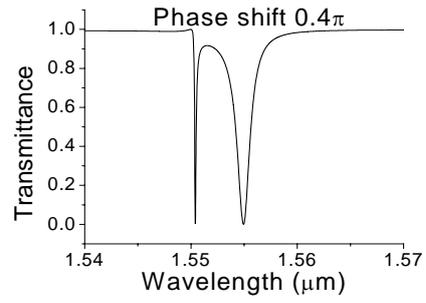
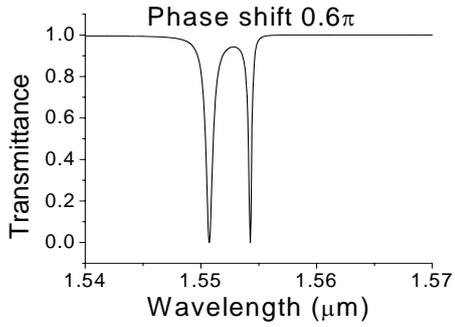
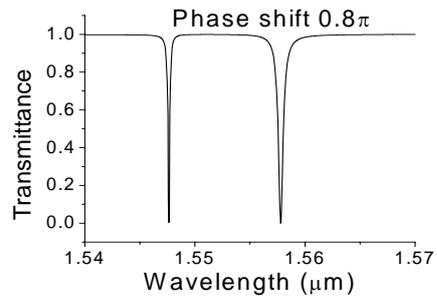
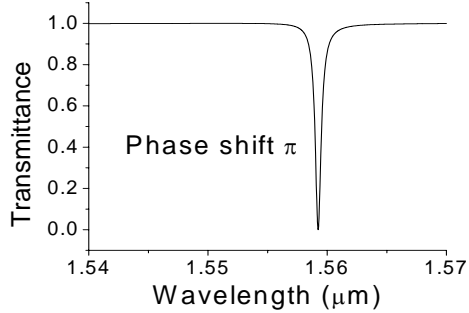

Fig.3

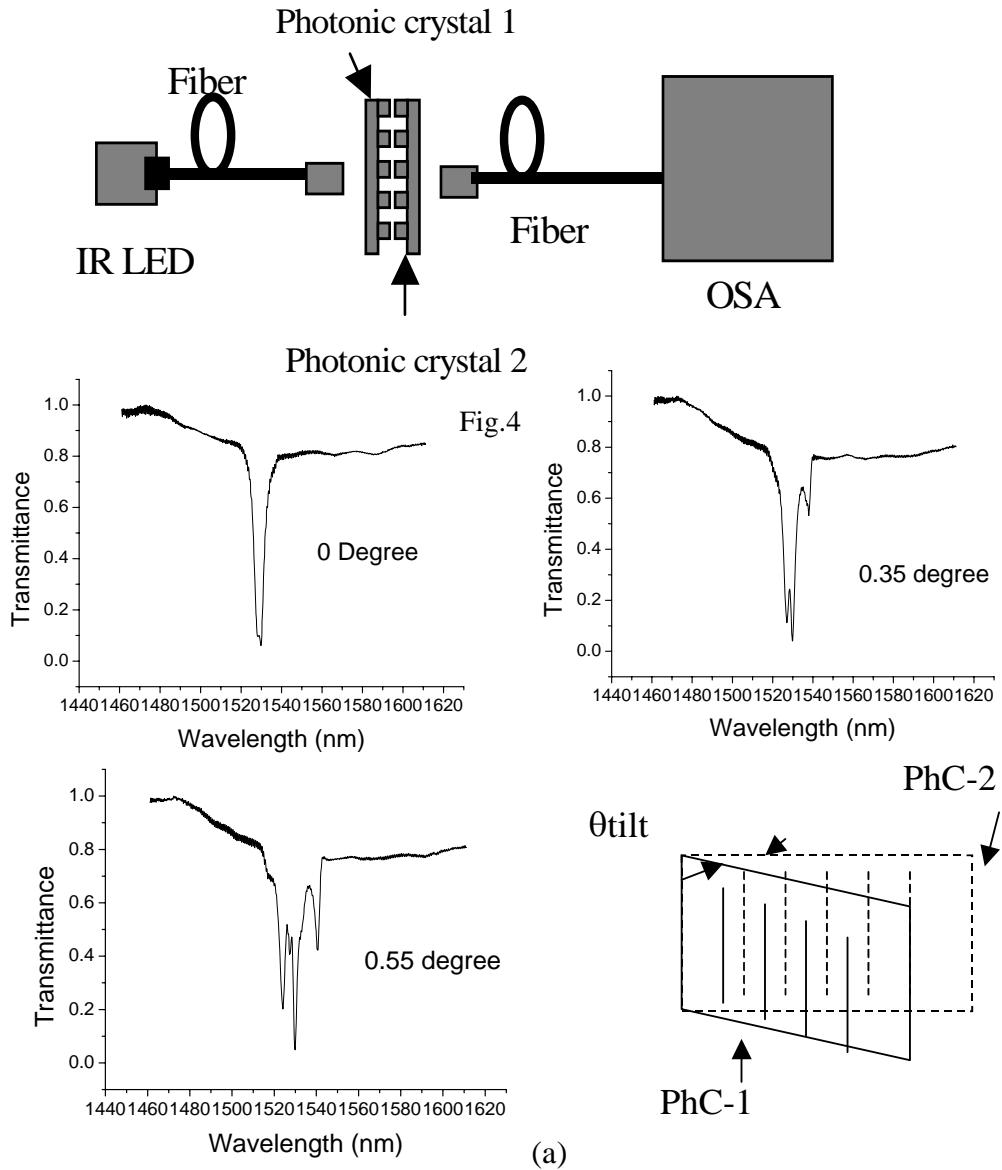

Fig.4

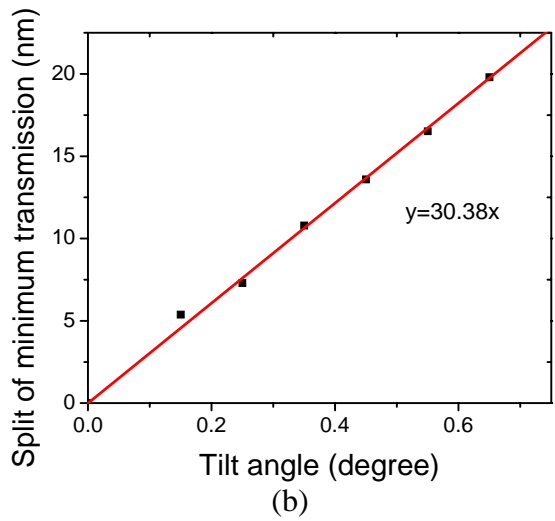

Fig.5

(a)

(b)

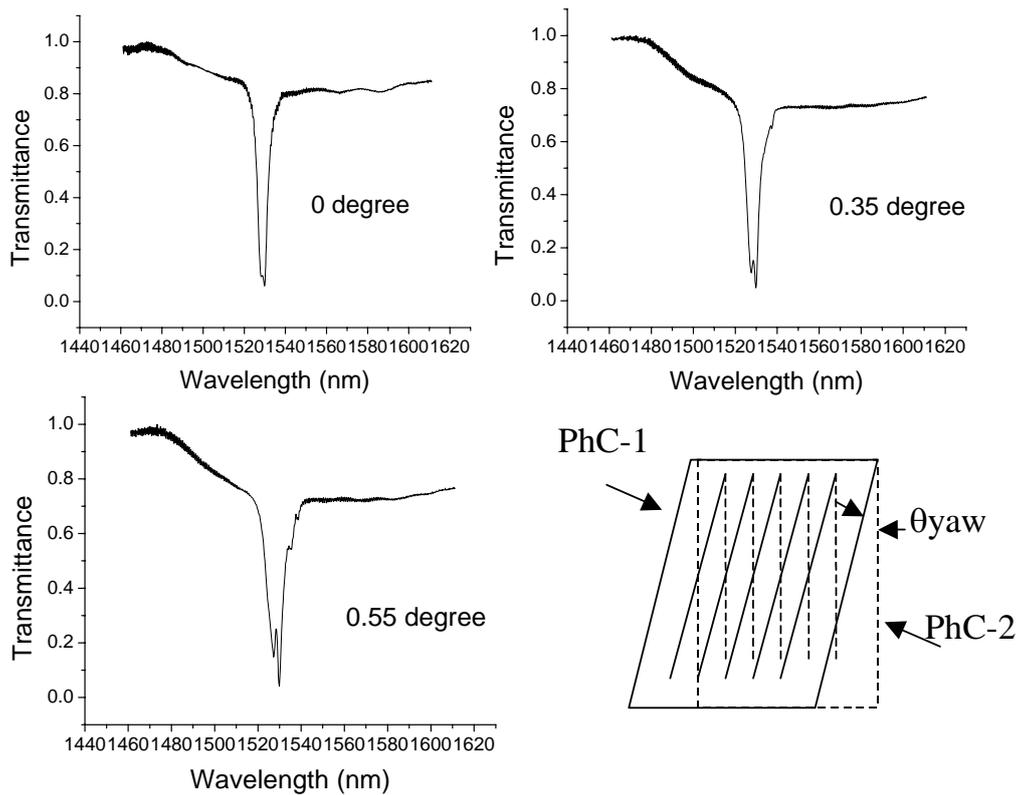

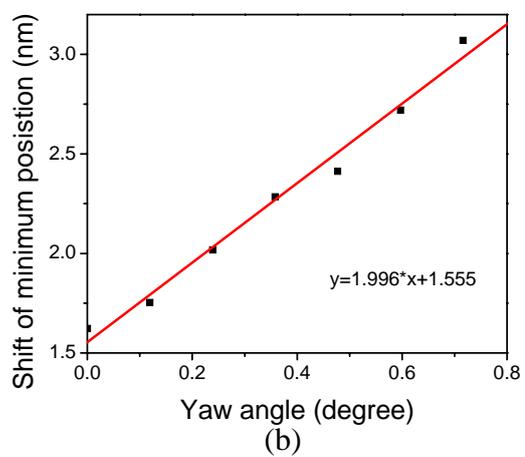

Fig.6